\documentclass{article}
\usepackage[margin=1in]{geometry}

\usepackage{mathtools}
\usepackage{booktabs,multirow,makecell}
\usepackage[flushleft]{threeparttable}
\usepackage{xcolor}

\usepackage{graphicx}

\usepackage{authblk}

\newcommand{\rev}[1]{\textcolor{black}{#1}}

\usepackage[numbers]{natbib}

\title{Modeling Discrete Coating Degradation Events via Hawkes Processes}

\author{Matthew Repasky$^1$}
\author{Henry Yuchi$^1$}
\author{Fritz Friedersdorf$^2$}
\author{Yao Xie$^1$}
\affil{
\small
$^1$H. Milton Stewart School of Industrial \& Systems Engineering, Georgia Institute of Technology \\
$^2$Luna Labs USA, Corrosion and Materials Performance}

\date{%
}

\begin{document}

\maketitle

\begin{abstract}
    
Forecasting the degradation of coated materials has long been a topic of critical interest in engineering, as it has enormous implications for both system maintenance and sustainable material use. Material degradation is affected by many factors, including the history of corrosion and characteristics of the environment, which can be measured by high-frequency sensors. However, the high volume of data produced by such sensors can inhibit efficient modeling and prediction. To alleviate this issue, we propose novel metrics for representing material degradation, taking the form of discrete degradation events. These events maintain the statistical properties of continuous sensor readings, such as correlation with time to coating failure and coefficient of variation at failure, but are composed of orders of magnitude fewer measurements. To forecast future degradation of the coating system, a marked Hawkes process models the events. We use the forecast of degradation to predict a future time of failure, exhibiting superior performance to the approach based on direct modeling of galvanic corrosion using continuous sensor measurements. While such maintenance is typically done on a regular basis, degradation models can enable informed condition-based maintenance, reducing unnecessary excess maintenance and preventing unexpected failures.

\end{abstract}


\section{Introduction}
Understanding the degradation of a coating system is essential to the sustainable use of coated materials. Coating systems are designed to inhibit corrosion of the underlying material, and periodic maintenance/re-application is necessary to avoid damage to coated panels. However, excessive re-application can be costly and hazardous. Coating system maintenance has significant implications for the environment and worker safety. Their synthesis often results in the release of hazardous air pollutants and volatile organic compounds, which can lead to issues such as health hazards and groundwater pollution~\cite{athappan2014emissions,reynolds1997preliminary,bell2006exposure}. Material corrosion costs the US Department of the Navy an estimated \$8.6 billion annually, limiting access to critical Department of Defense assets~\cite{shipilovimpact,auditofnavy}. Despite these concerns, maintenance is conducted in a fixed-time-interval manner, possibly leading to unnecessary maintenance which would entail unnecessary costs, material waste, and safety risks~\cite{auditofnavy}. To enable condition-based maintenance of coating systems, it is necessary to forecast their degradation given the present condition.

\begin{figure}[t]
    \centering
    \includegraphics[width=\textwidth]{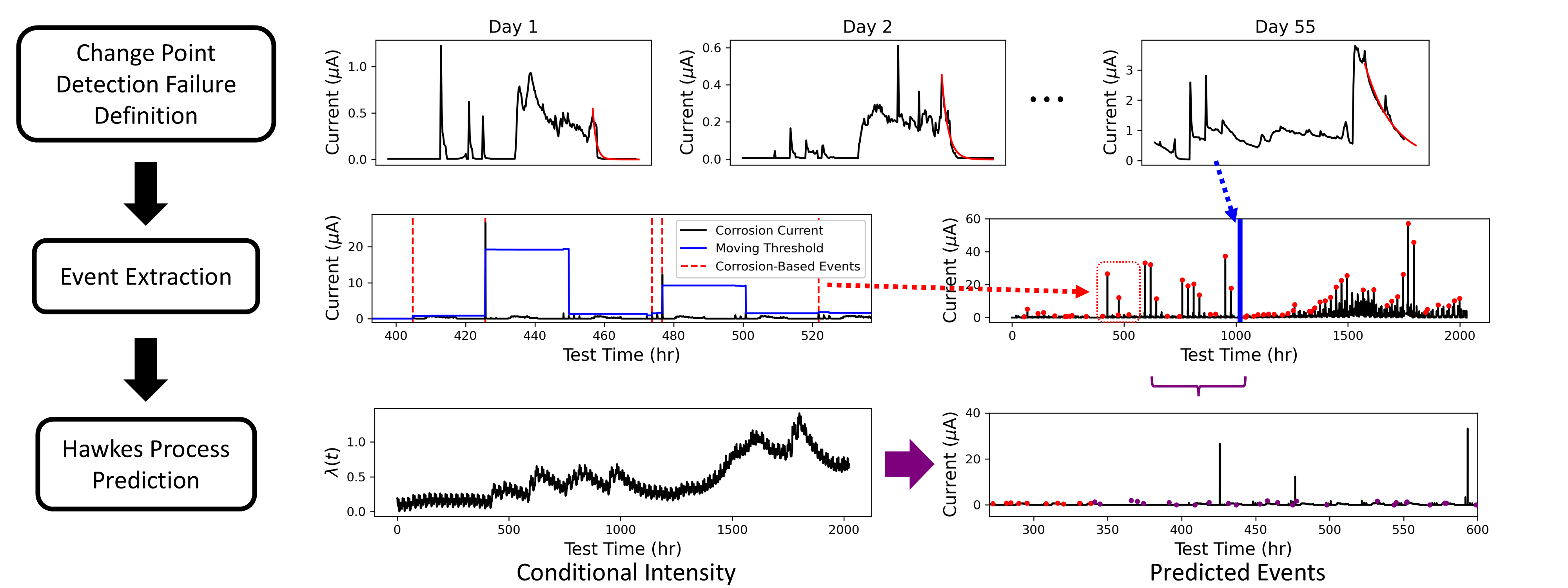}
    \caption{{\textbf{Flow Chart of Data Analysis. }\label{fig:flow_chart}}{The flow chart highlights the three main aspects of the modeling pipeline. First, failure is defined using a change point detection procedure measuring a change in the exponential tail in corrosion current data. Then, events corresponding to degradation are extracted from the continuous sensor readings. Finally, a point process model is fit to predict future degradation events.}}
\end{figure}

The degradation of coated materials can be due to corrosion, which is the result of spontaneous electrochemical reactions occurring on the surface, producing an electron flow (current) through the metal. The corrosion process can occur whenever sufficient humidity and contaminants are present, to provide a conducting electrolyte on the surface. The contact of two dissimilar materials can accelerate the corrosion process by creating a galvanic couple. Leveraging time series sensor measurements of galvanic corrosion current, relative humidity (RH), and surface conductance (indicating the presence of contaminants) poses sufficient information to facilitate the modeling of coating degradation, which we investigate in this work. To this end, Acuity zero-resistance ammeter (ZRA) sensors are used in all experiments.

Prediction of failure (e.g., blistering or cracking) can be accomplished using models of material degradation. Sensor measurements of the galvanic corrosion current can be used to build statistical models representing coating degradation. These can leverage environmental parameters and the history of degradation to forecast future corrosion. However, with typical measurement frequencies on the order of minutes, continuous raw sensor measurements provide an abundance of data on the time scales relevant to the corrosion process. Moreover, they may provide more information than necessary to represent degradation, resulting in potentially less efficient modeling. See Section~\ref{sec:analyze_events} for a thorough discussion and analysis. Representing degradation as a sequence of (relatively few) ``discrete events'' could facilitate more efficient modeling, and statistical models of discrete event data could then be applied to forecast degradation and predict failure.

Temporal point processes (TPPs) are a class of stochastic models used to predict discrete events in time~\cite{reinhart2018review}. Self-exciting TPPs (such as Hawkes processes) assume that past events impact the probability of future events~\cite{hawkes1971spectra}, which is represented by the conditional intensity. Extensions incorporate additional information (``marks'') in the conditional intensity. TPPs have already been widely used across the sciences, including epidemiology~\cite{chiang2022hawkes}, reliability analysis~\cite{zhang2020survival}, and seismology~\cite{kumazawa2014nonstationary}. A primary target when predicting failure of physical systems is characterizing the remaining time to failure. Stochastic approaches such as point processes~\cite{zhang2020survival} and Weibull analysis~\cite{pereira2018mechanical} are often applied. In this work, we aim to model and predict discrete degradation events in continuous time, providing a forecast for the degradation of a coated panel. Inspired by the literature characterizing corrosion as potentially self-exciting~\cite{mikhailov2006cooperative}, we model these events using a Hawkes process.

This work focuses on predicting failure of a coating system by forecasting its degradation. First, we define ``coating system failure'' in a data-driven manner, joining visual-inspection-based blister characterization and sensor-based change point detection (CPD). Then, we aim to define the discrete events used as metrics in this work, denoted the ``degradation events,'' and we demonstrate that these metrics are directly related to failure. Finally, a Hawkes process model is used to forecast degradation events for a given coated panel, which is used to predict failure. The contributions of our work are as follows:
\begin{itemize}
    \item We introduce a \textbf{novel definition of coating system failure}, defined in a data-driven manner using a CPD procedure based upon the characteristics of the galvanic corrosion current. This data-driven labeling of failure reconciles discrepancies in visual inspection of blistering due to differences in coating stack-up, which leads to a more physically-consistent failure definition across different experiments and materials.
    
    \item A flexible procedure is developed to extract \textbf{discrete degradation events from continuous sensor data}. These events capture salient features contributing to the degradation of a coating system, maintaining correlation with time to failure. The sparsity of events facilitates more efficient modeling of degradation. Rather than building a model to predict Galvanic corrosion directly, statistical models of degradation events are required to forecast orders of magnitude fewer measurements.
    
    \item Provided an initial period of observed degradation events, we use a \textbf{Hawkes process to forecast future event sequences}. This enables the characterization of the remaining protective life of a coating system. Specifically, for a given coated panel, we predict a future time interval at which failure may occur with high probability, which is based on the forecast of future degradation events.
\end{itemize}

\subsection{Related works}
Modeling the effect of galvanic corrosion current and environmental factors is supported by the understanding of corrosion from the perspective of materials science. High humidity in the presence of salt contaminants facilitates the electrochemical corrosion reactions~\cite{dehri2000effect,li2010eis}; the temperature has also been found to affect corrosion \cite{oje2019effect,sahir2014effect}. It has also been shown that corrosion blistering events may excite the onset of additional blistering \cite{mikhailov2006cooperative}, with other works accounting for the increasing permeability of coating layers as they deteriorate using non-homogeneous Poisson processes \cite{nicolai2007comparison}. Our approach builds upon these understandings of self-excitation and physical contributors to corrosion, applying a self-exciting point process model to predict material degradation in coating systems.

Sequential CPD procedures~\cite{xie2021sequential} are commonly applied in industrial applications, such as coating system failure and material degradation, to capture the point in time at which the system fails. A form of the cumulative sum (CUSUM) procedure~\cite{page1954continuous} is often applied, for instance differentiating normal process changes to more abrupt changes signifying failure in wheel coating~\cite{tout2018non}. Similarly, \cite{guo2021nonparametric} aim to detect deterioration in manufacturing processes to enable preventative maintenance using nonparametric distributional assumptions and trend detection in CPD. With respect to material degradation, \cite{theristis2021comparative} conducts a comparative analysis of CPD approaches applied to nonlinear degradation rates of photovoltaics, and \cite{chen2021two} develops degradation models which account for multiple changes in degradation characteristics. In this work, CPD is applied to ZRA sensor data measuring the instantaneous rate of corrosion, which experiences an observable change in diurnal pattern that can be captured to define a data-driven failure label.

\subsection{Overview}
The remainder of the paper is structured as follows. In Section~\ref{sec:data_overview}, the coated panel corrosion data are described, including a description of degradation experiments in laboratory and outdoor settings in Section~\ref{sec:data_description} and a definition of failure labels in Section~\ref{sec:failure}. In Section~\ref{sec:discrete_events}, multiple definitions of discrete events are defined using sensor measurement data, aiming to represent substantial contributions to the degradation of the coated panel. The relationship between these discrete events and failure/blistering are analyzed in Section~\ref{sec:analyze_events}. In Section~\ref{sec:model}, a Hawkes process is defined to model degradation events, which can be used to predict failure. Finally, experiments assessing the viability of the point process approach are highlighted in Section~\ref{sec:experiment}, followed by a discussion in Section~\ref{sec:discussion}.

\begin{figure}[t]
\centering
    {\includegraphics[width=\textwidth]{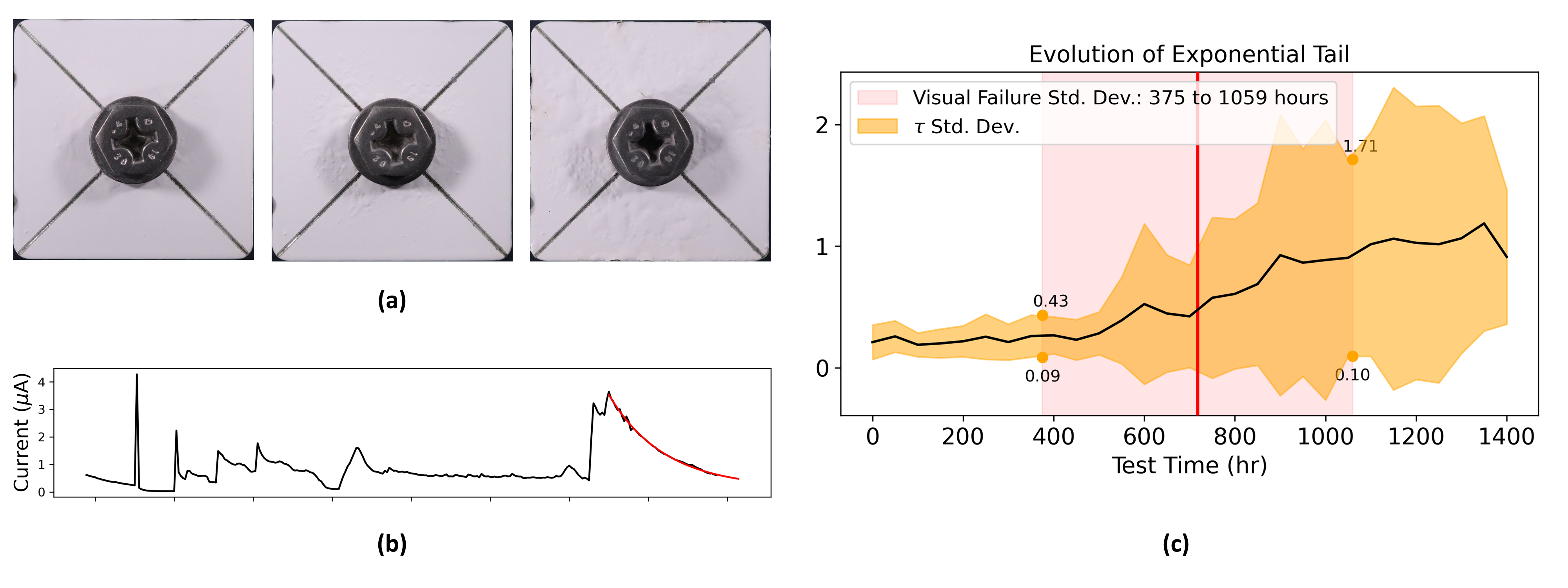}}
    \caption{\textbf{Progressive Blistering of Coated Panels. \label{fig:failure_definition}}{Figure (a) demonstrates progressive blistering of a coated panel over the course of an exposure experiment. In (b), the exponential tail of the diurnal cycle of galvanic corrosion current is demonstrated. (c) demonstrates that the exponential tail tends to increase over the course of all experiments, with a rapid increase concentrated during the periods in which visible blistering is observed.}}
\end{figure}

\section{Coating Degradation Sensor Data}\label{sec:data_overview}

In this section, we outline the datasets used in this study, which are derived from continuous sensor measurements of coated panels. We also define the failure mode, which is a data-driven label based on observed blistering on the surface of the coated panel, suggesting the degradation in the coating material.

\subsection{Experiments in laboratory and outdoors}\label{sec:data_description}
We analyze ZRA sensor data corresponding to the degradation of coated panels in four sets of experiment data. Three multi-sensor panels (MSPs) and one temperature sensor panel (TSP) are embedded on a single Luna Labs Acuity CR\textsuperscript{TM} docking platform system. The TSP and docking platform measure the surface temperature, air temperature, and relative humidity. Additionally, the TSP contains a gold interdigitated electrode (IDE) collecting 25 kHz conductance measurements, acting as a proxy for the measurement of contaminants on the material surface~\cite{ellicks2015measurements}. In this work, the data measured by the TSP 25 kHz gold IDE is generally called the ``conductance'' of the platform. The sensors of each MSP are coated, and include a 25 kHz gold IDE as in the TSP, a 10 kHz gold IDE, and a 7075 aluminium alloy and stainless steel 316 IDE. The measurements from the 7075/SS316 IDEs represent the galvanic corrosion rate (also called the galvanic corrosion current) of the underlying material via ZRA.

Three sets of the experiment data measured panels subject to a laboratory environment via an accelerated cyclic corrosion chamber: GMW 14872~\cite{gmw14872}. This chamber simulates accelerated environmental conditions via a repeated 24-hour cycle including 4 salt sprays over an 8-hour period, an 8-hour period of high humidity and salinity, and an 8-hour period of high temperature with low humidity. The first set of laboratory experiment data consists of four docking platforms (designated platforms 85-88) and is treated as described in the previous paragraph, which we denote the ACEM laboratory experiment data. The second set, consisting of two platforms (107 \& 113) and denoted the Scribe Width Comparison (SWC) data, compares the width of the scribe. The third set, also consisting of two platforms (135 \& 136), is denoted the Cracks versus Scribes (CVS) data and compares a crack-like defect to scribes. Due to the similarities in these three datasets, the three laboratory sets are pooled into a single dataset for most analyses. The final set of experiment data, consisting of 2 platforms (142 \& 143), represent panels which were exposed to outdoor conditions at the Battelle test site in Daytona Beach, Florida starting in March 2023. In the three laboratory datasets, the sensors report readings every 5 minutes, while the sensors have measurement period of 30 minutes in the outdoor experiments.

In each experiment, each MSP is wiped clean, deoxidized using Henkel Alumiprep 33, and pre-treated with Bonderite T5900 or Pre-Kote. Each MSP is then coated with one of four primers, including solvent-borne chromate and non-chromate aerospace primers (PPG-Deft CA7233 and PPG-Deft 02GN084) and water-borne chromate and non-chromate primers (Hentzen AD9318 and PPG-Deft 44GN098). Finally, one of two polyurethane topcoats is applied to each MSP (PPG 99GY001 and PPG-Deft CA8202). The cured coatings are scribed with an ``x'' shape before testing. See Table~\ref{tab:coating_stackups} for a description of the coating stackup of each sensor for each experiment.

\begin{table}[t]
\centering
\resizebox{\textwidth}{!}{
  {\begin{tabular}{r|cccccccccccccccccccc}
    \hline
    Pre-Treatment   & Bonderite*        & Bonderite         & Bonderite         & Bonderite         & Pre-Kote  & Pre-Kote  \\
    Primer          & CA7233            & AD9318            & 02GN084           & 44GN098           & 44GN098   & CA7233    \\
    Topcoat         & 99GY001           & 99GY001           & 99GY001           & 99GY001           & CA8202    & CA8202     \\
    \hline
    \multirowcell{4}{(Platforms, MSPs)} & \multirowcell{4}{(85,1)\\(87,3)\\(88,2)\\(143,1-3)} & \multirowcell{4}{(86,3)\\(87,1)\\(88,1)} & \multirowcell{4}{(85,3)\\(86,1)\\(88,3)} & \multirowcell{4}{(85,2)\\(86,2)\\(87,2)\\(142,1-3)} & \multirowcell{4}{(107,1-3)\\(113,1-3)\\(135,1-3)} & \multirowcell{4}{(136,1-3)} & & & \\ \\ \\  \\
    \hline
  \end{tabular}}
}
  \small{*Bonderite T5900}
  \caption{{Coating Stackups.\label{tab:coating_stackups}}}
\end{table}

\subsection{Failure mode}\label{sec:failure}
We aim to approximate the point at which small, discrete blisters begin forming across the surface of the coated panel, which we denote the \textit{failure time}. More specifically, we assume that the rate of blistering experiences a sudden change at some point in time, which could be related to cracking or the depletion of corrosion inhibitors in the primer. This change point is selected using two approaches in this work. The first approach uses direct visual observation of blister formation; failure is associated with each sensor roughly corresponding to the time at which small, sporadic blisters begin to appear (see Section~\ref{sec:blister_based_failure}). Failure can also be defined in a data-driven manner, where the goal is to capture the underlying physical change point, which is described in Section~\ref{sec:data_driven_failure}. \rev{The application of CPD for failure definition is demonstrated in the top panel of Figure~\ref{fig:flow_chart}.}

\subsubsection{Blister-based failure labels}\label{sec:blister_based_failure}
Visual inspection of the coated sensors reveals the onset of blistering over the course of exposure experiments. Figure~\ref{fig:failure_definition}a demonstrates the visible onset of blistering over the course of an experiment (from left to right). This observation acts as a direct measurement of the rate of blistering, but visual inspections are conducted once every 1-4 weeks. Therefore, these labels provide information at a much lower frequency than that of the sensor observations. Moreover, visual inspection is limited by observer bias and differences in experimental setup. The coating stackup applied to the sensors in the ACEM dataset utilizes a different topcoat than that used in the SWC and CVS datasets. In the ACEM data, the visual-inspection-based failure time is 1,030 hours on average (where the standard deviation is $\pm$223 hours), while the average failure time is 460$\pm$168 hours in the remaining two datasets. This difference in visually-observed blistering in otherwise similar experiments could be attributed to the difference in topcoat. That is, if two coated panels have different topcoats, it may be more or less difficult to visually observe blisters, despite the fact that the materials might have experienced the same amount of degradation.

\begin{figure}[t]
\centering
    {\includegraphics[width=\textwidth]{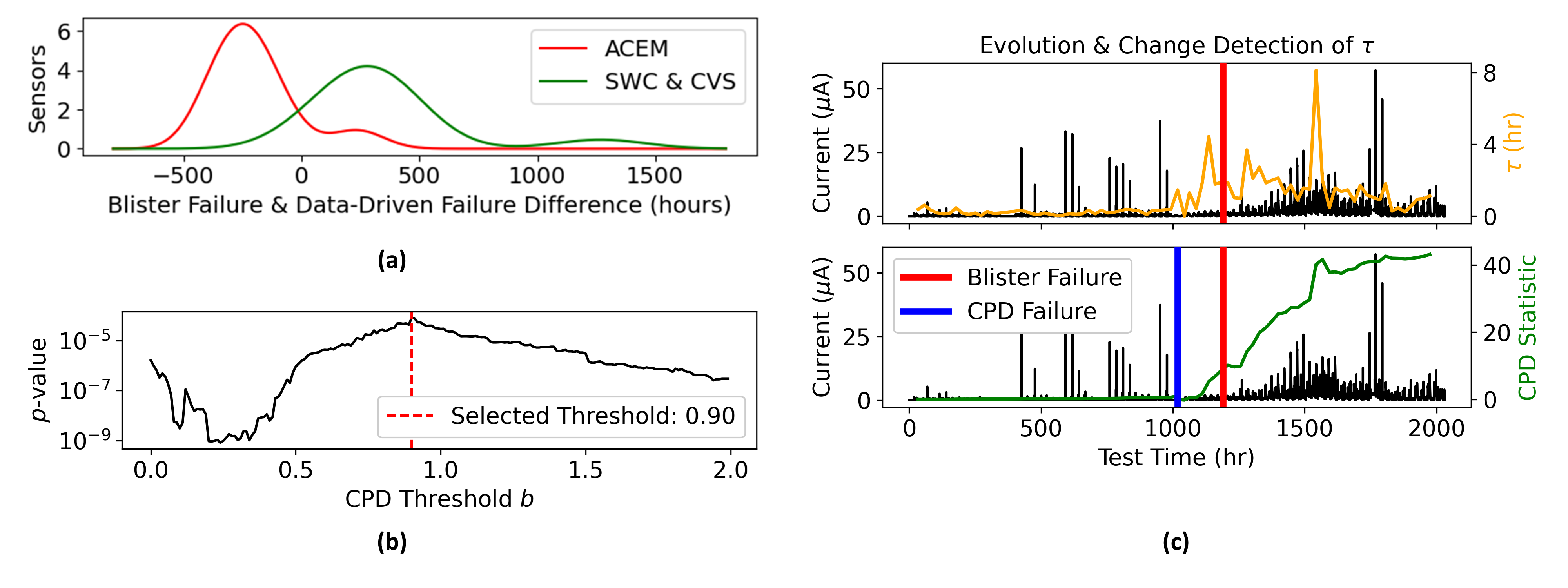}}
    \caption{\textbf{Data-Driven Failure Detection. \label{fig:failure_t_test}}{In (a), the $p$-values of T-tests between the difference in blister-based and data-driven failure times between the ACEM and the SWC \& CVS datasets are visualized for a range of CPD thresholds. In (b), the resultant two distributions which maximize the $p$-value are shown, corresponding to $\hat{b}=0.90$. In (c), the evolution of $\tau$ is shown in the top panel, while the bottom panel demonstrates the WLCUSUM change point detection for data-driven failure labeling.}}
\end{figure}

\begin{figure}[t]
    \centering
    {\includegraphics[width=0.6\textwidth]{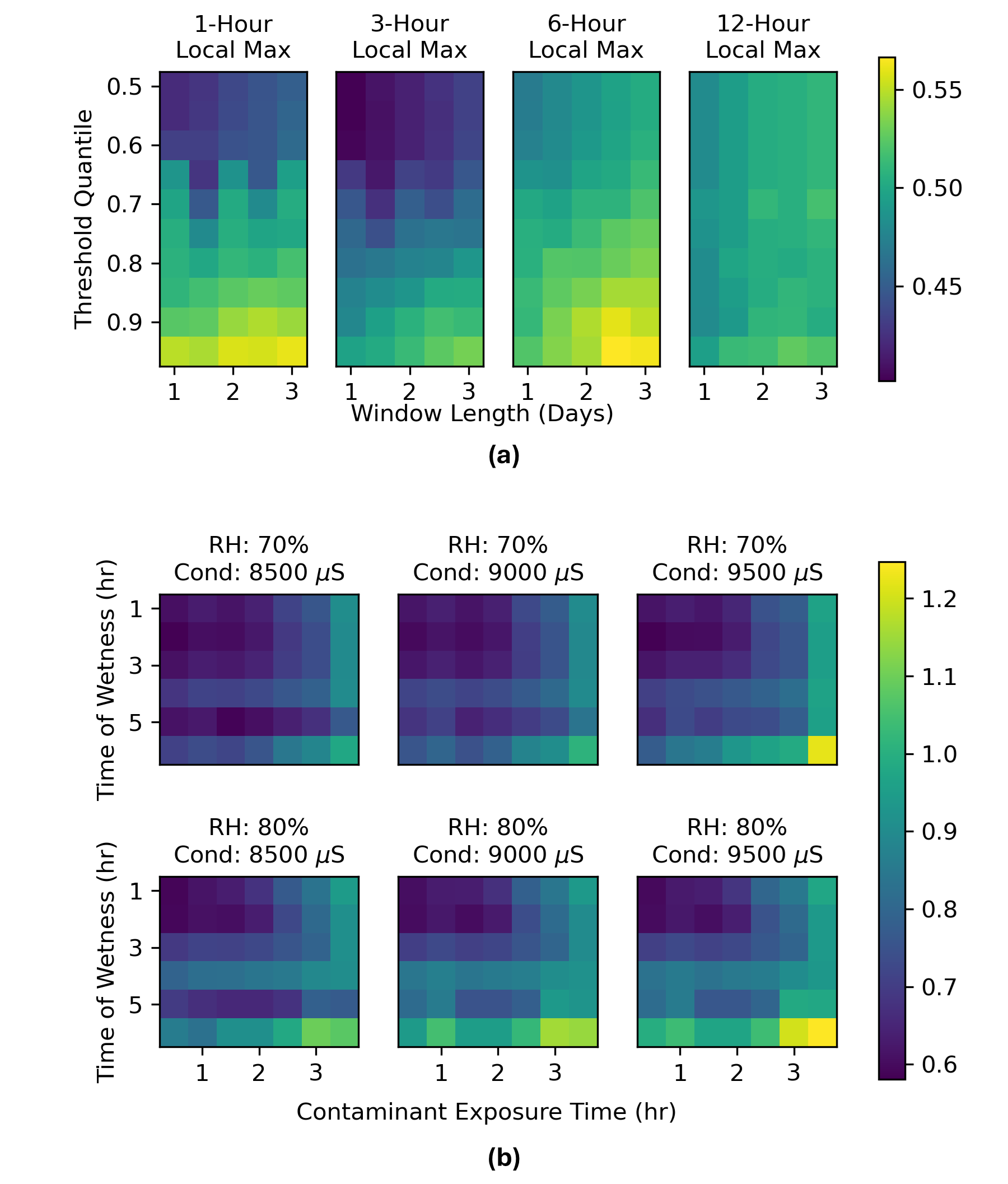}}
    \caption{\textbf{Event Parameter Selection. \label{fig:event_param_grid}}{In (a), the coefficient of variation (CV) of the distribution of the number of events at the time of failure is visualized for each parameter combination defining corrosion events. Similarly, (b) represents a grid search over environment event definition parameters.}}
\end{figure}

\subsubsection{Data-driven failure labels}\label{sec:data_driven_failure}
Failure time can also be assigned in a data-driven manner, which can act to reconcile discrepancies arising due to differences in coating stack-up across experiments. Additionally, the higher frequency of the sensor data measurements (with respect to the visual observations) provides more fine-grained information about the status of the coating. It is observed that the galvanic corrosion current has a recurring pattern corresponding to the diurnal cycles in the experiment data. A key feature of this cycle is an exponential tail at the end of each cycle, which may be fit to the following functional form:
\begin{equation*}
    y(t) = A \cdot \exp\left(-\frac{t-t_0}{\tau}\right) + B.
\end{equation*}

See Figure~\ref{fig:failure_definition}b for a demonstration of the exponential model fit of a diurnal cycle (see the top panel of Figure~\ref{fig:failure_t_test}c for the evolution of $\tau$ in time for an individual sensor). We observe that the time constant $\tau$ begins to grow as blisters form along the coating surface. Considering the average corrosion current and the average value of $\tau$ in the laboratory data over binned time (50-hour bins) over a set of calibration experiments, Spearman's rank correlation coefficient reveals that $\tau$ is positively correlated (0.661) with corrosion current. In Figure~\ref{fig:failure_definition}c, the visual inspection failure time labels are visualized via the standard deviation of their distribution. The time-evolving distribution of $\tau$ is visualized by the black line (mean) and its standard deviation (orange). The increasing trend, on average, of $\tau$, is concentrated near the center of the visual inspection failure time distribution, indicating that there exists a relationship between visually-observed blisters and the data-driven variable $\tau$. Therefore, a positive shift in $\tau$ can be associated with the onset of blistering.

\rev{To automatically detect failure and label its time, we use sequential CPD. This technique allows for the detection of} distributional changes in time series observations in a data-driven manner~\cite{xie2021sequential}. The assumption is that data follow pre-change distribution $f_0$ for some time before experiencing a change at time $\gamma$, after which the data follow post-change distribution $f_1$. Generally, CPD procedures define a statistic $S_i$ indexed by $i=0,1,\dots$ over sequential observations $X_i$. The CUSUM procedure~\cite{page1954continuous} uses the cumulative log-likelihood ratio of a sequence:
\begin{equation*}
    S_i = \sum_{k=1}^i \ell(X_k),
\end{equation*}
where the log-likelihood ratio $\ell(X)\coloneqq\log f_1(X)/f_0(X)$. Notably, $\ell$ is negative before the change and positive after, so $S_i$ has negative drift before the change and positive drift after.

The CUSUM procedure detects a change when $W_i\coloneqq S_i-\text{min}_{k\leq i} S_k$, the amount $S_i$ exceeds its past minimum, exceeds some threshold $b>0$. It can be shown that this may be computed recursively~\cite{xie2021sequential}:
\begin{equation*}
    W_i = \text{max}(W_{i-1},0) + \ell(X_i), \quad\quad W_0=0.
\end{equation*}
To avoid distributional assumptions on $f_1$, it can be estimated via $f_\theta$, resulting in $\ell_\theta(X)\coloneqq\log f_\theta(X)/f_0(X)$ in the generalized likelihood ratio (GLR) approach~\cite{lorden1971procedures,lorden1973open}. The distribution $f_\theta$ can be estimated using a windowing procedure, and the corresponding CPD procedure is known as the window-limited CUSUM (WLCUSUM), which also permits recursive computation~\cite{xie2023window}:
\begin{equation*}
    W^L_i = \text{max}(W^L_{i-1},0) + \ell_{\hat{\theta}}(X_i), \quad\quad W^L_0=0.
\end{equation*}
In WLCUSUM, $\hat{\theta}$ is an estimate of $\theta$ generated using a length-$w$ window of data, denoted $X_{i-w:i}$, via maximum likelihood estimation (MLE), for instance. The estimated change $\hat{\gamma}$ is also declared when $W^L_i$ first exceeds some threshold $b>0$.

A WLCUSUM procedure is conducted to detect the change in $\tau$,
\begin{equation*}
    W^L_i = \text{max}(W^L_{i-1},0) + \bar{\tau}_{i-w:i} \left( \tau_i - \frac{\bar{\tau}_{i-w:i}}{2} \right),
\end{equation*}
with $W^L_0 = 0$, The assumption is that $\tau_i$ are Gaussian-distributed and experience a positive shift in mean after the change point, which is computed using the $w$-window average value of $\tau$, denoted $\bar{\tau}_{i-w:i}$. The CPD threshold is tuned to minimize the discrepancy in failure definition between two sets of experiments. Specifically, since the goal is identify a failure definition which best reconciles the differences between the ACEM data and the SWC \& CVS data, the difference between the blister-based failure time label and the $\tau$-based failure time label is computed for each sensor for a range of CPD thresholds $b$. The selected threshold $\hat{b}=0.90$ maximizes the $p$-value of a T-test between the distribution of failure time differences of the ACEM data and the SWC \& CVS data, which is visualized in Figure~\ref{fig:failure_t_test}a and (b). The intuition is that the difference between blister-based failure and data-driven failure should be indistinguishable between the two types of laboratory data. The resultant data-driven failure is visualized for a single sensor in Figure~\ref{fig:failure_t_test}c.

\begin{figure}[t]
\centering
    {\includegraphics[width=\textwidth]{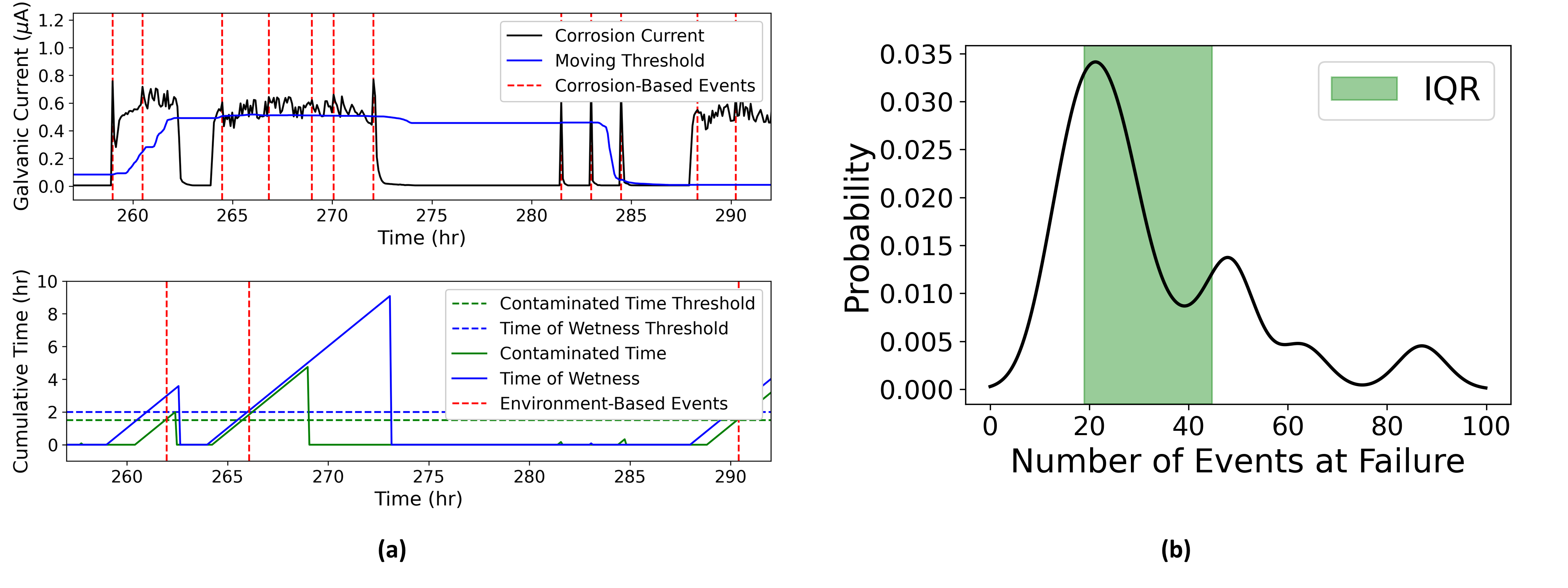}}
    \caption{\textbf{Discrete Events from Continuous Sensor Measurements. \label{fig:event_extraction}}{In (a), corrosion events (top panel) and environment events (bottom panel) are demonstrated. Corrosion events occur when local maxima in galvanic corrosion current rise above a moving threshold. Environment events occur when the RH and conductance are persistently above specified thresholds. In (b), the empirical distribution of the number of environment events at failure in laboratory data is visualized, exhibiting low variation.}}
\end{figure}

\section{Coating Degradation Events}\label{sec:discrete_events}
 
In this section, we outline definitions for discrete events capturing degradation, which are meant as sparse markers of corrosion. Fluctuations in the diurnal cycle as captured by conductance and RH measurements are modeled, suggesting that such features represent impactful corrosion patterns. This reduces the size of the data, facilitating efficient modeling. We define discrete degradation events from continuous-time data extracted from sensor readings. Each sensor has an associated failure time (see Section~\ref{sec:failure}); each failure is associated with a number of degradation events under some event definition. Using training data, either (1) an empirical distribution of the number of events at failure can be maintained or (2) a parametric distribution (e.g., a Gaussian distribution) of events at failure can be estimated using the observations in the training data. We outline three options for the definition of degradation events. The parameters of each event definition are tuned to minimize the variability of the number of events at failure (the coefficient of variation (CV) of the empirical distribution) in a calibration split of the laboratory datasets. This includes 18 of the 24 laboratory MSPs (7 chromate and 11 non-chromate, see Section~\ref{sec:lab_exp}). \rev{See Figure~\ref{fig:flow_chart} for a representation of the event extraction procedure in the context of the entire data analysis workflow.}

\subsection{Definition of discrete events}\label{sec:event_definitions}

\subsubsection{Corrosion events}\label{sec:corr_even_definition}
Peaks in galvanic corrosion current as measured by ZRA sensors can be extracted. The threshold in galvanic corrosion current to constitute aggressive corrosion (a peak) is computed via a moving-window quantile of the current. An event is a local maximum above the moving threshold. There are three parameters: (i) the window length, (ii) the quantile of the moving threshold, and (iii) the size of the local period to define local maxima. Each event is defined by the time and magnitude of the peak galvanic corrosion current. A grid search over these three parameters is conducted on the calibration spit of the data, and the CV of the distribution of the number of events at failure time is visualized for each parameter combination in Figure~\ref{fig:event_param_grid}a. The minimum-CV corrosion events are defined as $\pm$3-hour local maxima above a 1-day moving window of the 0.55-quantile in galvanic corrosion current. A visualization of the events can be seen in Fig~\ref{fig:event_extraction}a (top).

\subsubsection{Environment events}
Since corrosion is facilitated by high RH and contaminants, the time of wetness (duration that RH is above a threshold) and contaminant exposure time (similarly defined for the 25 kHz gold conductance) can capture hazardous environmental conditions. An environment event occurs when the time of wetness and contaminant exposure time are both above a threshold. There are four parameters: the thresholds in (i) RH, (ii) conductance, (iii) time of wetness, and (iv) contaminant exposure time. Events are defined by the earliest time both metrics are above their thresholds without dropping below the thresholds, see Figure~\ref{fig:event_extraction}a (bottom). Conducting a grid search over the four parameters on the calibration data (Figure~\ref{fig:event_param_grid}b), the minimum-CV environment events use an RH threshold of 70\% for more than 2 hours and a conductance threshold of 9,500 $\mu$S for more than 0.5 hours.

\subsubsection{Hybrid events}
Corrosion events and environment events can be combined to define a hybrid type of degradation event. Specifically, an event occurs when galvanic corrosion current is locally maximal during wet and contaminated conditions. The seven parameters include (i) corrosion window length, (ii) corrosion quantile, (iii) the local period definition, (iv) the RH threshold, (v) the conductance threshold, (vi) the time of wetness, and (vii) the contaminant exposure time. Due to the exponential increase in the size of the parameter space with respect to the number of parameters, the grid search selection was conducted over environment parameters (iv)-(vii), with the corrosion event parameters (i)-(iii) fixed to the values defined in Section~\ref{sec:corr_even_definition}. The resultant events occur when a corrosion current reading is a $\pm$3-hour local maximum above a 1-day window of the 0.55-quantile corrosion current, while simultaneously the RH is over 70\% for more than 4 hours and the conductance is over 7,000 $\mu$S for more than 0.5 hours.

\begin{table}[t]
\centering
\resizebox{\textwidth}{!}{
  {\begin{tabular}{r|cccc}
    \hline
    \multirowcell{2}{Metric} & \multirowcell{2}{Mean Measurements\\at Failure} & \multirowcell{2}{Coefficient\\of Variation} & \multirowcell{2}{Time to Failure\\Pearson Correlation} & \multirowcell{2}{Time to Failure\\Spearman Correlation} \\
     \\
    \hline
    Corrosion Events & 89 & 0.40 & -0.46 & -0.50 \\
    Environment Events & 33 & 0.58 & -0.37 & -0.40 \\
    Hybrid Events & 18 & 0.72 & -0.45 & -0.44 \\
    Charge & 9700 & 0.40 & -0.54 & -0.62  \\
    \hline
  \end{tabular}}
}
  \caption{Analysis of each metric with respect to failure time in the calibration data. Correlations are significant with $p<0.01$.\label{tab:event_analysis}}
\end{table}

\begin{figure}[t]
\centering
    {\includegraphics[width=0.5\textwidth]{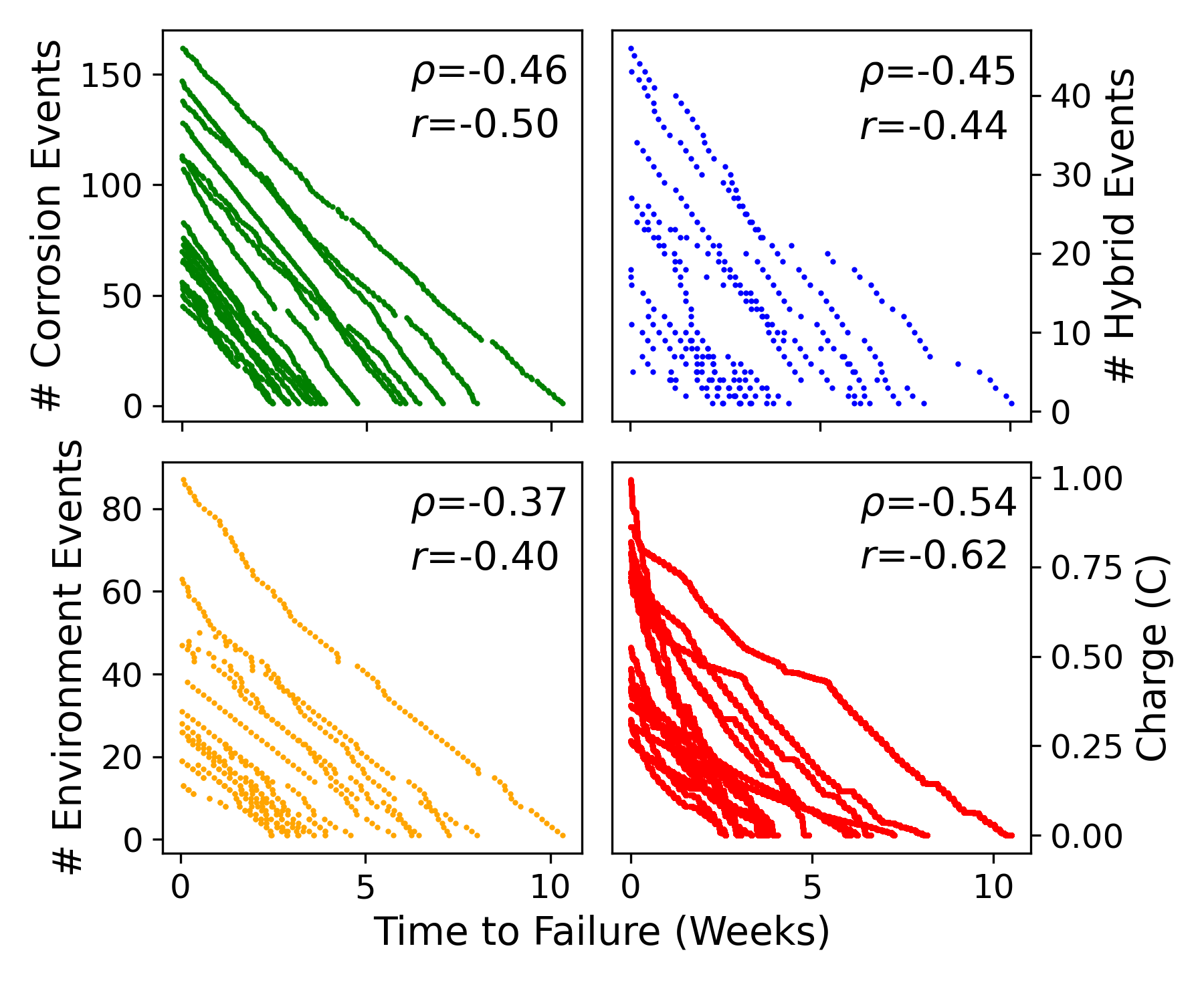}}
    \caption{\textbf{Discrete Events versus Accumulated Charge. \label{fig:event_v_ttf}}{The accumulation of degradation metrics (discrete events or continuous charge) are plotted against time to failure. In each sub-plot, the correlation between the accumulated metric and the time to failure is reported.}}
\end{figure}

\subsection{Analysis of discrete events}\label{sec:analyze_events}
Discrete degradation events provide a potentially more salient metric for representing the corrosion process than direct measurements of the galvanic corrosion current. In other words, we suggest that degradation events are nearly as effective as corrosion current in terms of predicting failure, but require far fewer measurements. This can be accomplished by tracking the accumulation of events and the accumulation of charge (via the time-integrated corrosion current) to predict the time of failure. In this section, we analyze the properties of the number of events at failure in the calibration split of the data (18 sensors, see Section~\ref{sec:lab_exp}), compared with the accumulated charge at failure time. In the first column of Table~\ref{tab:event_analysis}, it can be seen that predicting failure based on accumulated events requires forecasting of 2-3 orders of magnitude fewer measurements than accumulated charge. The variability in each metric can be assessed by comparing the CV at failure time, which is displayed in the second column of Table~\ref{tab:event_analysis}. Corrosion events and accumulated charge both have the lowest CV of 0.40. This indicates a tighter concentration of the metric at failure, potentially making failure easier to predict given an accurate model of corrosion events or corrosion current. Finally, the (Pearson's $\rho$ and Spearman's $r$) correlation between each metric and the time to failure is displayed in the final two columns of Table~\ref{tab:event_analysis}. The event accumulation metrics (especially corrosion events) exhibit similar correlation with respect to the time to failure as the accumulated charge. See Figure~\ref{fig:event_v_ttf} for a visualization of the accumulation of each degradation metric plotted against the time to failure.

The CV and correlation analyses reveal that discrete degradation events capture much of the relationship between corrosion current and failure. In particular, the corrosion-based discrete events are the most highly-correlated events with respect to time to failure, and maintain the same CV as accumulated charge. Moreover, given sensor measurements every 5 minutes, the number of events at failure is over 100 times smaller than the total number of sensor measurements. That is, degradation events are far more sparse without sacrificing the predictive properties of corrosion current. From a modeling perspective, in order to predict failure using degradation events, a model is only required to accurately forecast $O(10)$ events, as opposed to the forecasting of $O(1000)$ measurements required by a model of corrosion current. In the following sections, we outline a statistical model to forecast discrete degradation events (Section~\ref{sec:model}), and empirically demonstrate the superior capability of discrete-event-based models compared with models of corrosion current (Section~\ref{sec:experiment}).

\begin{figure}[t]
\centering
    {\includegraphics[width=0.6\linewidth]{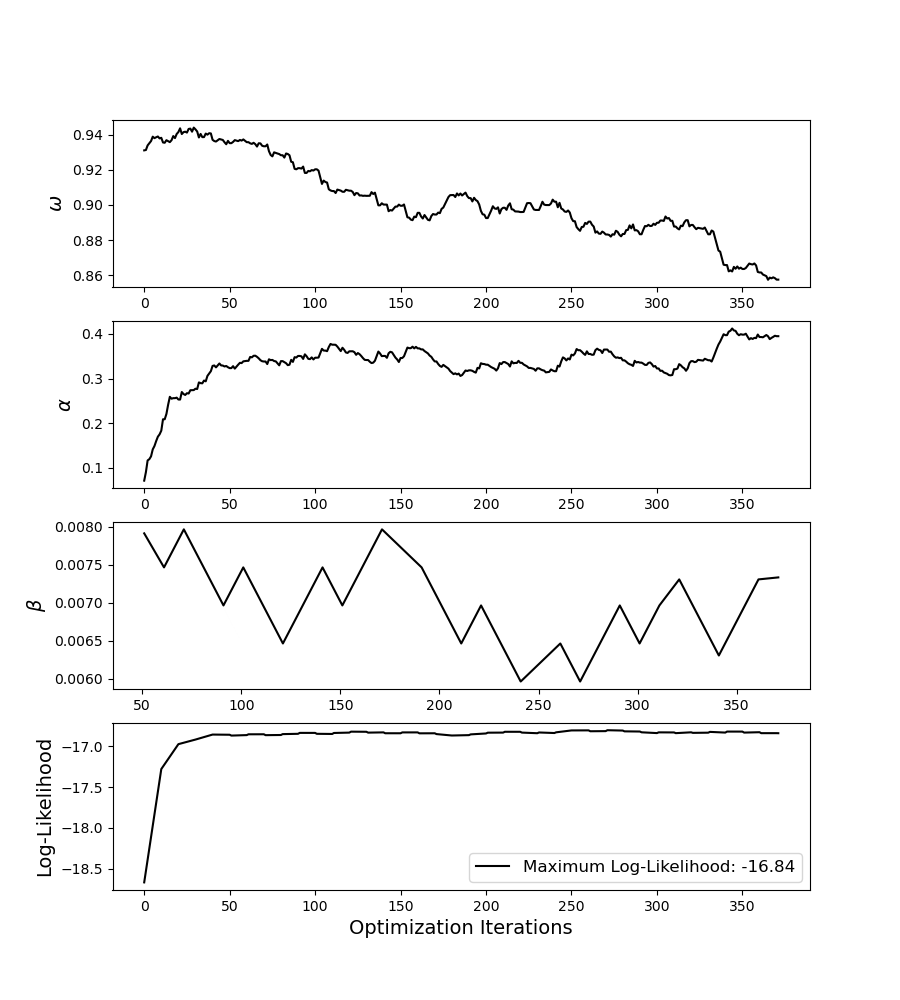}}
    \caption{\textbf{Hawkes Process Optimization Dynamic. \label{fig:hybrid_optimization}}{The optimization dynamic of the laboratory experiment hybrid event Hawkes process of Section~\ref{sec:lab_exp}. The evolution of the parameters $\alpha$, $\omega$, and $\beta$ are also displayed. The method converges after about 375 iterations of gradient descent.}}
\end{figure}

\section{Modeling Degradation Events}\label{sec:model}
A forecast of future degradation can be used to predict failure time. In Section~\ref{sec:discrete_events}, we outlined novel metrics which quantify degradation as sparse discrete events in continuous time, and we demonstrated that these metrics are correlated with failure in Section~\ref{sec:analyze_events}. Given an initial period of observed degradation events, it is therefore useful to construct a forecast of future degradation to predict failure time. TPPs provide a natural statistical framework for modeling discrete events in continuous time. In Section~\ref{sec:degradation_hawkes_process}, we outline a particular TPP (a Hawkes process) for modeling degradation events. This forecast can be used to predict failure, as described in Section~\ref{sec:failure_prediction}. \rev{Figure~\ref{fig:flow_chart} shows how the fitting and prediction via TPP modeling fits into the entire anlysis pipeline.}

\subsection{Marked Hawkes process}
TPPs are stochastic processes~\cite{daley2003introduction} which model the occurrence of discrete events in continuous time. A marked TPP is described by its intensity $\lambda$, which captures the probability of an events (at times $t$ with associated information known as \textit{marks} $m$) occurring in an infinitesimal time interval. Inhomogeneous, marked TPPs model $\lambda(t,m)$ as a time evolving function that may depend on conditional information such as the history of events $\mathcal{H}_t=\{(t_i,m_i):t_i<t\}$:
\begin{equation}\label{eq:full_hawkes}
    \lambda(t,m|\mathcal{H}_t) = \lambda_g(t|\mathcal{H}_t)f(m|t,\mathcal{H}_t),
\end{equation}
where $\lambda_g$ is called the ground process conditional temporal intensity and $f$ is the conditional mark density. The Hawkes process model~\cite{hawkes1971spectra,reinhart2018review} is a form of inhomogeneous TPP which assumes events are self-exciting, such that the history of events  affects $\lambda_g$ via a non-negative kernel $k$:
\begin{equation}\label{eq:general_hawkes_model}
    \lambda_g(t|\mathcal{H}_t) = \lambda_0(t) + \sum_{\{i:t_i\in\mathcal{H}_t\}} g(m_i) \cdot k(t-t_i).
\end{equation}
We assume the history of marks $m_i$ affect $\lambda_g$ via fixed, positive function $g$. In general, $\lambda_0$, $g$, and $k$ may be parameterized by some set of parameters $\Theta$. \rev{In Section~\ref{sec:degradation_hawkes_process}, we provide a specific form and parameterization of the Hawkes process model for degradation modeling.}

Provided a trajectory of $n$ events over time horizon $T$, the log-likelihood function of $\lambda_g$ has the following form~\cite{daley2003introduction}:
\begin{equation}\label{eq:loglik}
    \ell(\Theta) = \sum_{i=1}^n \log(\lambda_g(t_i)) - \int_0^T \lambda_g(t)dt.
\end{equation}
Maximum likelihood estimation (MLE) of ground process parameters $\Theta$ is usually analytically intractable. For small datasets, numerical maximization of Equation~\ref{eq:loglik} is often conducted instead~\cite{reinhart2018review}, including numerical computation of the integral. \rev{In the implementation used in this work, the integral is computed with the midpoint rule using time discretization of 1 hour.}

\begin{figure}[t]
\centering
    {\includegraphics[width=0.5\textwidth]{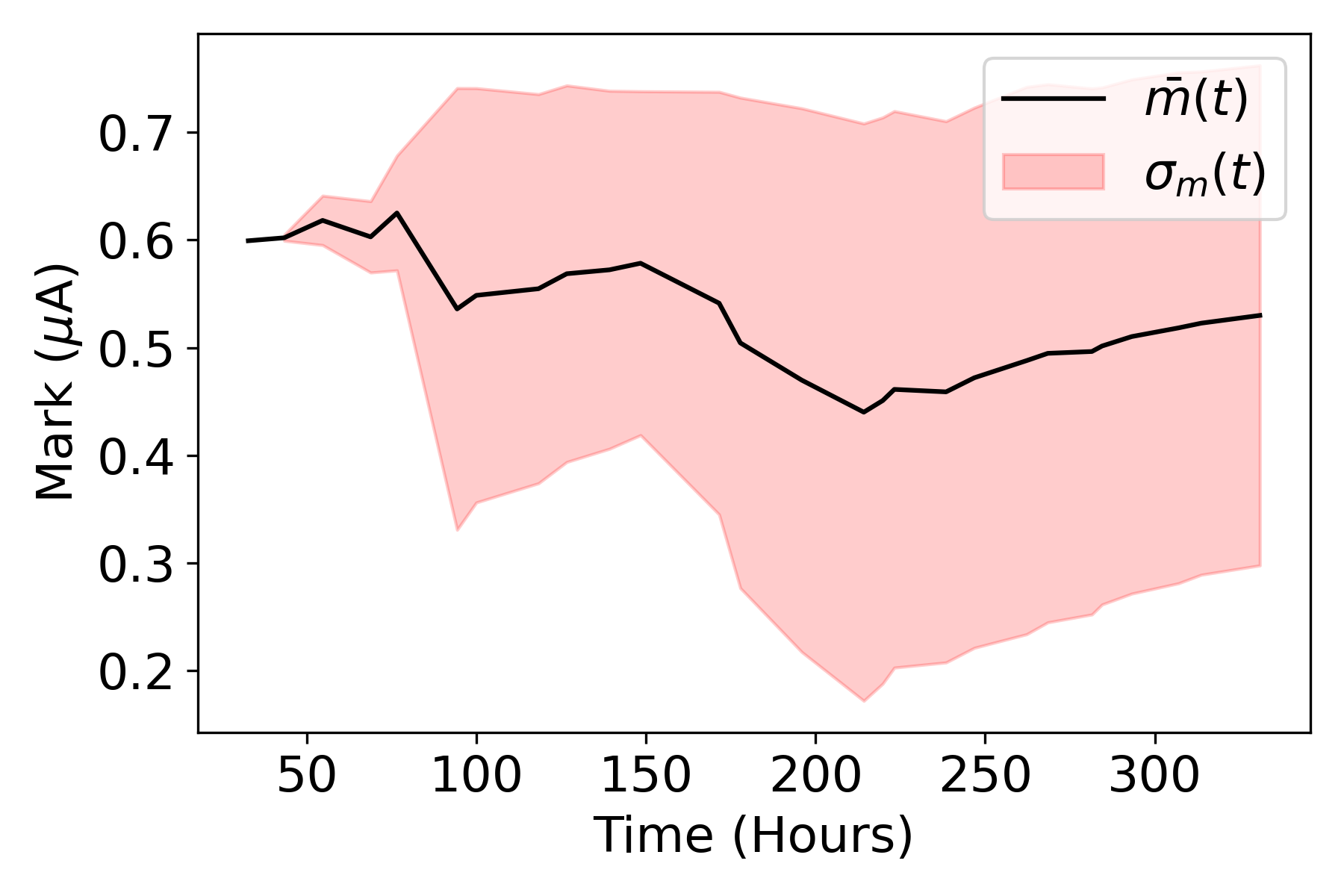}}
    \caption{{\textbf{Time-Varying Estimate of Mean and Standard Deviation of Mark. }\label{fig:fit_marks}}{\rev{The parameters of the condition mark density (Equation~\ref{eq:conditional_mark_density}), i.e., the time-evolving mean and standard deviation of mark, are shown for a sequence of corrosion degradation events.}}}
\end{figure}

\subsubsection{Degradation event Hawkes process}\label{sec:degradation_hawkes_process}
In this work, we use a Gaussian to model the conditional mark density \rev{in Equation~\ref{eq:full_hawkes}}:
\begin{equation}\label{eq:conditional_mark_density}
    f(m|t,\mathcal{H}_t) = \mathcal{N}\left( \bar{m}(t), \sigma_m^2(t) \right),
\end{equation}
where $\bar{m}(t)$ and $\sigma_m^2(t)$ are the mean and variance of all marks $m$ up to time $t$ (i.e., $\{ m_i,  t_i\in\mathcal{H}_t \}$). \rev{See Figure~\ref{fig:fit_marks} for a demonstration of the fitted mean and standard deviation of marks function for a time interval in a training data sensor (see Section~\ref{sec:experiment} for details on training data).} We use the following form for the ground process intensity to model degradation events\rev{, given data $\mathcal{H}_t=\{(t_i,m_i):t_i<t\}$}:
\begin{equation}\label{eq:our_hawkes_model}
    \lambda_g(t | \mathcal{H}_t) = \alpha \mu(t) + \omega\sum_{\{i : t_i \in \mathcal{H}_t\}} m_i \beta \exp{(-\beta(t-t_i))}.
\end{equation}
\rev{Therefore, relating to each term in the general form of Equation~\ref{eq:general_hawkes_model}, in our model (Equation~\ref{eq:our_hawkes_model}) the background intensity $\lambda_0(t)=\alpha\mu(t)$ is a periodic (daily) kernel density estimation (KDE), $g(m_i)=m_i$, and $k(t-t_i)=\beta\exp{(-\beta(t-t_i))}$. The parameters to be estimated via MLE of Equation~\ref{eq:loglik} are $\Theta=\{\alpha, \omega, \beta\}$.} Coefficients $\alpha$ and $\omega$ capture the relative impact of the background and self-excitation. Coefficient $\beta$ is the self-excitation influence decay.

Due to the small size of the data considered in this work, parameters $\Theta:=\{\alpha,\beta,\omega\}$ are estimated using numerical MLE with respect to Equation~\ref{eq:loglik}. The parameters are initialized using metrics obtained via a DBSCAN clustering of events in time. Kernel weight $\omega$ is initialized to the fraction of isolated events that initiate a cluster. Self-excitation decay $\beta$ is initialized as the reciprocal of the average cluster duration. Background intensity weight $\alpha$ is initialized to the overall rate of events in time. Gradient descent is conducted on the negative log-likelihood (NLL) with respect to $\alpha$ and $\omega$ followed by NLL line search in $\beta$, repeating this alternating approach until convergence. The gradient descent is conducted using the stochastic gradient descent (SGD) optimizer in PyTorch. The coordinate descent optimization is conducted as follows. (1) Parameters $\alpha$ and $\beta$ are updated via SGD on randomly sampled sequences from the training set for each iteration. The NLL on the entire training dataset is computed every 10 iterations of gradient descent. (2) When the relative difference in subsequent values of the NLL falls below a threshold, the parameter $\beta$ is updated via line search on a random training trajectory. If the subsequent NLL relative difference is below a threshold (with respect to the most recent gradient descent iteration), then optimization terminates. Otherwise, the process repeats from step (1).

\subsubsection{Treatment of each event type}
The goal is to model each event type defined in Section~\ref{sec:event_definitions} using (three separate) marked Hawkes process models as defined in Section~\ref{sec:degradation_hawkes_process}. For all three degradation event types, the time $t$ of an event is simply the time at which the event occurs. For corrosion events and hybrid events, the mark $m$ is defined as the value of the corrosion current (in $\mu$A) at the time of the event. Since environment events do not depend on the corrosion current, we treat environment events as unmarked, which is equivalent to setting $m=1$ for each environment event.

\subsection{Failure Prediction}\label{sec:failure_prediction}

After learning the parameters of the Hawkes process, an initial trajectory of events (e.g., the first few weeks of measurements) can be provided to forecast future events via Poisson thinning~\cite{daley2003introduction}. This algorithm involves sampling from a homogeneous Poisson process initially, followed by selectively retaining a subset of events according to the inhomogeneous conditional intensity. To generate more robust predictions, several ($n_{\rm traj}$) event trajectories are sampled per sensor. After event trajectories are forecast for a coated panel, predicted events are associated with a prediction for the time to failure. The estimated distribution of degradation events at the time of failure can be used to produce a high-probability-of-failure future time interval based on events predicted using the Hawkes process.

Given an observed distribution of events at failure time as described in Section~\ref{sec:discrete_events}, a time window predicting high probability failure can be constructed by predicting the number of events $n_q$ associated with two $q$-quantiles of this distribution. We choose the $q=0.25$ and $q=0.75$ quantiles such that the prediction window is associated with the inter-quartile range (IQR) of the distribution (we consider separate distributions for chromate- and non-chromate-inhibited coated panels). The point process can be used to estimate the expected time at which the requisite number of events, $n_{0.25}$ and $n_{0.75}$, has occurred. This is done using the $n_{\rm traj}$ number of event trajectories sampled for the sensor from the Hawkes process; the average time at which $n_{0.25}$ and $n_{0.75}$ number of events occurred is recorded.

\begin{table}[t]
\centering
  {\begin{tabular}{r|cccccc}
    \hline
    & \multicolumn{3}{c}{Laboratory} & \multicolumn{3}{c}{Outdoor} \\
     \cmidrule(lr){2-4} \cmidrule(lr){5-7}
    Method & $n_{\rm train}$ & $n_{\rm eval}$ & $\bar{T}$ & $n_{\rm train}$ & $n_{\rm eval}$ & $\bar{T}$ \\
    \hline
    Corrosion Events     & 665 & 226 & 336 & 54 & 50 & 334 \\
    Environment Events   & 249 & 81  & 328 & - & - & - \\
    Hybrid Events        & 126 & 44  & 308 & 12 & 21 & 334 \\
    \hline
  \end{tabular}}
  \caption{Number of events and time horizon in experiments\label{tab:event_parameters}}
\end{table}

\begin{figure}[t]
\centering
    {\includegraphics[width=\textwidth]{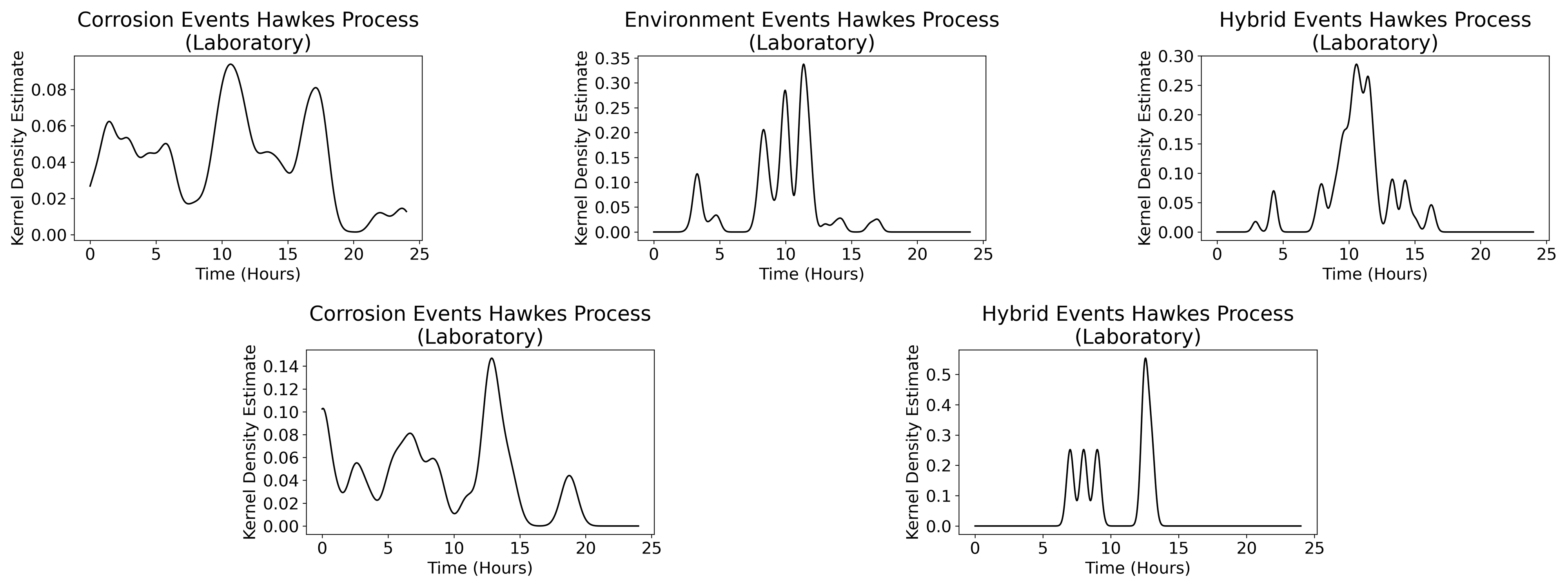}}
    \caption{{\textbf{Background Intensity Estimates. }\label{fig:kde_grid}}{\rev{Visualization of daily kernel density estimate (KDE) of background intensity $\mu(t)$ (Equation~\ref{eq:our_hawkes_model}) for each event definition in the laboratory and outdoor settings.}}}
\end{figure}

\section{Experiments}\label{sec:experiment}
The coating degradation event Hawkes process model is applied to modeling sensor readings from two experimental settings. In Section~\ref{sec:lab_exp}, a set of 24 coated sensors in a controlled-environment, laboratory setting are analyzed. In Section~\ref{sec:outdoor_exp}, 6 coated sensors are placed in an outdoor environment in Battelle Florida Materials Research Facility, subject to an uncontrolled environment setting.

For comparison, we utilize a baseline method that directly predicts the galvanic corrosion current to model degradation. A linear vector auto-regressive model of the galvanic corrosion current which incorporates continuous measurements of temperature, RH, 25 kHz Au conductance, and galvanic corrosion current is trained on the first two weeks of each training sensor. Ordinary least squares (OLS) models are evaluated, in addition to linear models regularized using $\ell_2$ (ridge) and $\ell_1$ (LASSO) regularization. When forecasting future measurements, the true environment conditions (temperature, RH, 25 kHz Au conductance) are provided at all future time steps. The forecast accumulation of charge via the time-integrated forecast corrosion current is used to construct high-probability-of-failure windows. These windows are the same as those described in Section~\ref{sec:failure_prediction}: accumulated-charge-based windows are constructed using the 0.25 and 0.75 quantiles of the distribution of accumulated charge at the time of failure in the calibration split, where separate thresholds are used for chromate- and non-chromate-inhibited coatings.

\begin{table}[t]
\centering
  {\begin{tabular}{r|cccccc}
    \hline
    & \multicolumn{3}{c}{Laboratory} & \multicolumn{3}{c}{Outdoor} \\
     \cmidrule(lr){2-4} \cmidrule(lr){5-7}
    Method & $\hat{\alpha}$ & $\hat{\omega}$ & $\hat{\beta} \ ({\rm hr}^{-1})$ & $\hat{\alpha}$ & $\hat{\omega}$ & $\hat{\beta} \ ({\rm hr}^{-1})$ \\
    \hline
    Corrosion Event Hawkes Process & 2.19 & 0.72 & $8.98\times10^{-3}$ & 0.59 & 0.82 & $2.24\times10^{-3}$ \\
    Environment Event Hawkes Process & 0.59 & 0.82 & $2.24\times10^{-3}$ & - & - & - \\
    Hybrid Event Hawkes Process & 0.39 & 0.86 & $7.33\times10^{-3}$ & 0.32 & 1.11 & $6.41\times10^{-3}$ \\
    \hline
  \end{tabular}}
  \caption{Estimated Hawkes process parameters setting\label{tab:fit_parameters}}
\end{table}

\begin{table}[t]
\centering
  {\begin{tabular}{r|cccc}
    \hline
    & \multicolumn{2}{c}{All Sensors} & \multicolumn{2}{c}{Test Only} \\
     \cmidrule(lr){2-3} \cmidrule(lr){4-5}
    \multirowcell{2}{\\Method} & \multirowcell{2}{Mean Width\\(weeks)} & \multirowcell{2}{Mean Error\\(weeks)} & \multirowcell{2}{Mean Width\\(weeks)} & \multirowcell{2}{Mean Error\\(weeks)} \\
     \\
    \hline
    Corrosion Event Hawkes Process & 2.22 & 1.32 & 2.72 & 1.80 \\
    Environment Event Hawkes Process & \textbf{1.57} & 1.26 & \textbf{1.53} & 1.32 \\
    Hybrid Event Hawkes Process & 2.61 & \textbf{1.11} & 2.75 & \textbf{0.75} \\
    \hline
    Ordinary Least Squares & 4.25 & 1.17 & 4.31 & 1.51 \\
    Ridge Regression & 4.02 & 1.31 & 3.72 & 2.74 \\
    LASSO Regression & 4.03 & 1.30 & 3.73 & 2.72 \\
    \hline
  \end{tabular}}
  \caption{Failure prediction window evaluation in the laboratory setting\label{tab:prediction_window_metrics_lab}}
\end{table}

\begin{figure}[t]
\centering
    {\includegraphics[width=\textwidth]{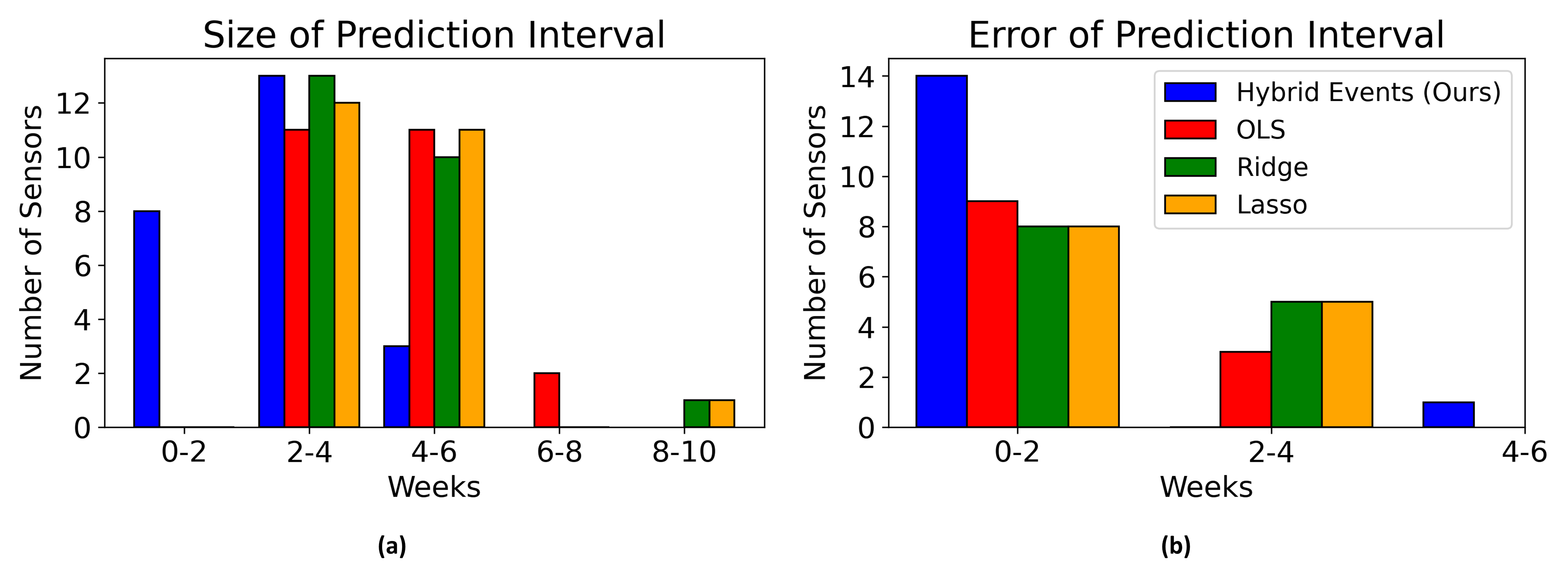}}
    \caption{\textbf{Laboratory Modeling Results. \label{fig:interval_stats_lab}}{Visualization of model performance on entire laboratory dataset. In (a), the interval sizes (in 2-week bins) are visualized for each method. In (b), the discrepancies when true failures are not captured by the prediction intervals are visualized.}}
\end{figure}

`
\subsection{Laboratory Experiments}\label{sec:lab_exp}
The eight panels subjected to the laboratory test (see Section~\ref{sec:data_description}) house three MSPs each, which record measurements every five minutes. Nine of these MSPs are coated with chromated primers (CA7233 \& AD9318) and 15 are coated with non-chromate primers (02GN084 \& 44GN098). Model training (MLE and interval quantile determination) is conducted using 18 of these sensors (seven chromate and 11 non-chromate); this is the same split used to calibrate event definitions in Section~\ref{sec:discrete_events}. The generalization of the approach is validated on the remaining six sensors (the test data).

\subsubsection{Failure Prediction} The goal is to predict the failure time as defined in Section~\ref{sec:data_driven_failure}, which is based on the metric $\tau$. The distribution of the number of events at the time of failure is constructed using the training data. Prediction in this section focuses on constructing high-probability-of-failure windows as described in Section~\ref{sec:failure_prediction} to predict the data-driven failure label.

\subsubsection{Event Definitions} Corrosion, environment, and hybrid events (see Section~\ref{sec:discrete_events}) are evaluated as predictors for failure. \rev{Table~\ref{tab:event_parameters} displays the total number of train events ($n_{\rm train}$) and evaluation events ($n_{\rm eval}$), in addition to the average time horizon $\hat{T}$ for MLE (Equation~\ref{eq:loglik}).} The distribution of the number of environment events at failure in the training data is shown in Figure~\ref{fig:event_extraction}b. The quartiles of these distributions are used as the target numbers of events for failure prediction. For each event type, two sets of values ($n_{0.25}$ and $n_{0.75}$) are chosen, one pair for each type of coating.

\subsubsection{Optimization Details}
\rev{The KDEs of the background intensity $\mu(t)$ (see Equation~\ref{eq:our_hawkes_model}) are shown for each event definition in Figure~\ref{fig:kde_grid}.} The MLE optimization of the Hawkes process parameters uses the first two weeks of event sequences in the 18 calibration sensors (14 for training, 4 for validation). We use the coordinate-descent-style procedure as described in Section~\ref{sec:degradation_hawkes_process}, with gradient clipping. Optimization terminates when the relative difference (between subsequent epochs) in validation NLL falls below a threshold. For optimizing linear auto-regressive baselines, the mean-squared error in the validation data is used to determine the lag size and regularization coefficients. The specifics for each Hawkes model are as follows.
\begin{itemize}
    \item \textbf{Corrosion Events Hawkes Process:} The $\alpha$ and $\omega$ gradients are clipped to $[0,10^3]$ with a learning rate of $5\times10^{-4}$. The $\beta$ line search is conducted over 20 equally-spaced points in a $\pm1\times10^{-2}$ radius surrounding the previous value of $\beta$. The optimization of $\alpha$ and $\omega$ terminates when the relative difference in validation NLL is less than $10^{-4}$. The algorithm terminates when the relative difference in validation NLL after line search is less than $10^{-4}$.

    \item \textbf{Environment Events Hawkes Process:} The $\alpha$ and $\omega$ gradients are clipped to $[0,10^3]$ with a learning rate of $1\times10^{-3}$. Line search is conducted over 20 points in a $\pm1\times10^{-3}$ vicinity. The $\alpha/\omega$ optimization terminates when the relative difference in validation NLL is less than $9\times10^{-5}$. Optimization terminates when the relative difference in validation NLL after line search is less than $9\times10^{-5}$.

    \item \textbf{Hybrid Events Hawkes Process:} The $\alpha$ and $\omega$ gradients are clipped to $[0,10^3]$ with a learning rate of $1\times10^{-3}$. The $\beta$ line search is conducted over 20 equally-spaced points in a $\pm5\times10^{-4}$ radius surrounding the previous value of $\beta$. The optimization of $\alpha$ and $\omega$ terminates when the relative difference in validation NLL is less than $10^{-4}$. The algorithm terminates when the relative difference in validation NLL after line search is less than $10^{-4}$. The maximum log-likelihood optimization trajectory for is shown in Figure~\ref{fig:hybrid_optimization}.
\end{itemize}
For each sensor, $n_{\rm traj}=10$ trajectories are sampled from the fit Hawkes process models. \rev{See Table~\ref{tab:fit_parameters} for the Hawkes process parameters fit using MLE.}

\subsubsection{Results} The results are summarized in Table~\ref{tab:prediction_window_metrics_lab}. The mean width reflects the duration of the prediction intervals, and the mean error reflects the average true failure time difference from the edge of the interval (for failures which fall outside the interval). The discrete event-based windows are generally less than 3 weeks wide, see Figure~\ref{fig:interval_stats_lab}a, with average error generally less than two weeks. Figure~\ref{fig:interval_stats_lab}b shows that nearly all failures are within two weeks of the predicted interval for discrete event methods. The linear auto-regressive methods fail to meaningful windows. While the average error of the baseline is similar to event-based methods (about 1-2 weeks), the windows are much larger (about a week larger in the test data). Figure~\ref{fig:interval_stats_lab} highlights the heavier tails of the window size distribution generated by the linear baselines.

\begin{table}[t]
\centering
  {\begin{tabular}{r|cccc}
    \hline
    & \multicolumn{2}{c}{All Sensors} & \multicolumn{2}{c}{Test Only} \\
     \cmidrule(lr){2-3} \cmidrule(lr){4-5}
    \multirowcell{2}{\\Method} & \multirowcell{2}{Mean Width\\(weeks)} & \multirowcell{2}{Mean Error\\(weeks)} & \multirowcell{2}{Mean Width\\(weeks)} & \multirowcell{2}{Mean Error\\(weeks)} \\
     \\
    \hline
    Corrosion Event Hawkes Process & 10.3 & \textbf{18.8} & 10.6 & \textbf{18.8} \\
    Hybrid Event Hawkes Process & 19.0 & 19.2 & 19.0 & 19.2  \\
    \hline
    Ordinary Least Squares & 10.6 & 23.9 & 9.6 & 35.1 \\
    Ridge Regression & \textbf{2.61} & 35.7 & \textbf{1.76} & 54.7  \\
    \hline
  \end{tabular}}
  \caption{Failure prediction window evaluation in the outdoor setting\label{tab:prediction_window_metrics_outdoor}}
\end{table}

\begin{figure}[t]
\centering
    {\includegraphics[width=\textwidth]{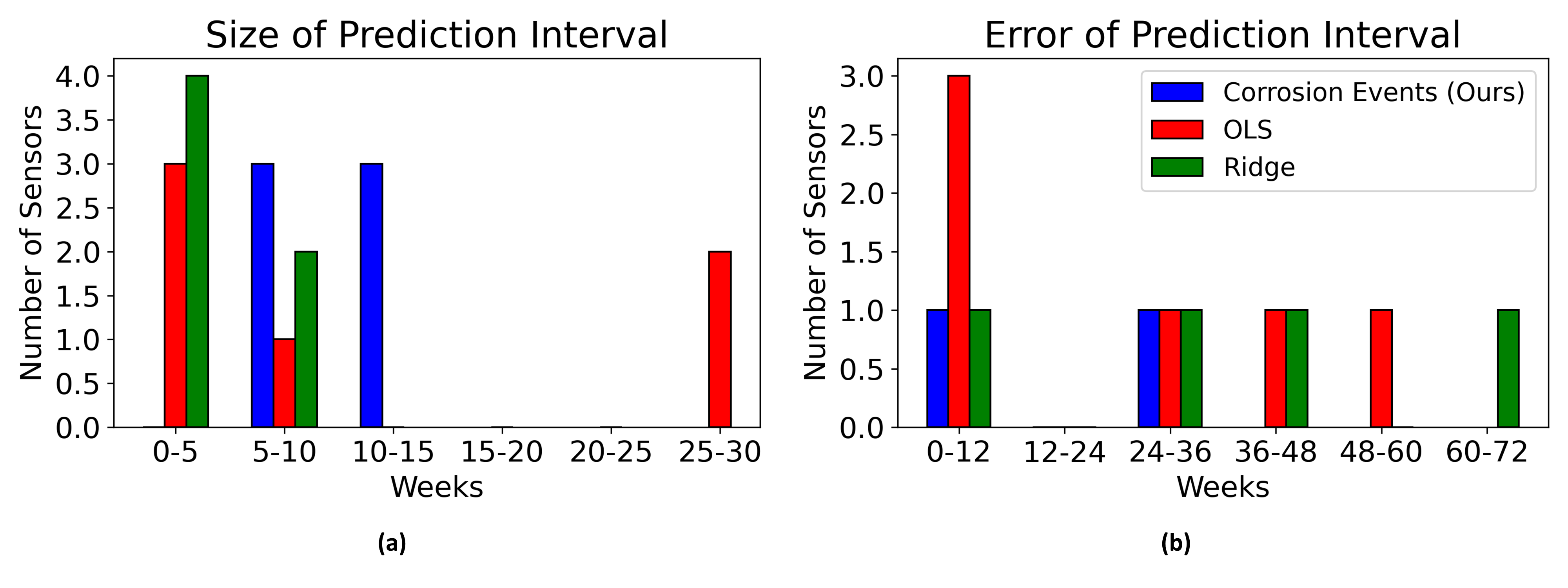}}
    \caption{\textbf{Outdoor Modeling Results. \label{fig:interval_stats_outdoor}}{Visualization of model performance on entire outdoor dataset. In (a), the interval sizes (in 5-week bins) are visualized for each method. In (b), the prediction errors (in 12-week bins) are visualized.}}
\end{figure}

\subsection{Outdoor Experiments}\label{sec:outdoor_exp}

The variable-environment outdoor dataset consists of two panels of three sensors that record measurements every 30 minutes (six total coated sensors). Half of the sensors are coated with chromate inhibitors while the other half are coated with non-chromate inhibitors. Models are calibrated using 3 of these sensors, 1 chromate and 2 non-chromate, with generalization evaluated on the remaining 3 sensors (the test data). The calibration partition is split into training data (1 chromate and 1 non-chromate sensor) and validation data.

\subsubsection{Failure Prediction} Unlike the laboratory experiments, all the data are generated from the same experiment. As such, the blister-based failure label (Section~\ref{sec:blister_based_failure}) can be expected to be consistent in a physically meaningful way, so it is used as the failure label in this analysis. Furthermore, given the relatively few number of sensors, it is not possible to generate an empirical distribution of the number of events at failure in these data. As such, a parametric (Gaussian) distribution is defined which is centered on the number of events at failure in the training sensor for each coating with CV equal to the CV of the empirical distributions of the laboratory data.

\subsubsection{Event Definitions} In this section, Hawkes processes trained using corrosion events and hybrid events are evaluated (Section~\ref{sec:discrete_events}). The corrosion event definition is directly translated from that of the laboratory data, while the hybrid event definition uses mostly the same parameters as the laboratory data, with the exception being that the conductance threshold is adjusted to 50 $\mu$S (due to an observed difference in the conductance pattern between the laboratory and environment). \rev{As in the laboratory data, Table~\ref{tab:event_parameters} shows the number of train events, evaluation events, and the average time horizon $\hat{T}$.}

\subsubsection{Optimization Details}
\rev{The fit background intensities for the outdoor data are also displayed in Figure~\ref{fig:kde_grid}.} As in Section~\ref{sec:lab_exp}, Hawkes process coordinate descent optimization is trained using the first two weeks in the 3 calibration sensors (2 for training, 1 for validation). The details of hyper-parameter choices are as follows:
\begin{itemize}
    \item \textbf{Corrosion Events Hawkes Process:} The $\alpha$ and $\omega$ gradients are clipped to $[0,10^3]$ with a learning rate of $5\times10^{-4}$. The $\beta$ line search is conducted over 20 equally-spaced points in a $\pm1\times10^{-2}$ radius surrounding the previous value of $\beta$. The optimization of $\alpha$ and $\omega$ terminates when the relative difference in validation NLL is less than $10^{-4}$. The algorithm terminates when the relative difference in validation NLL after line search is less than $10^{-4}$.

    \item \textbf{Hybrid Events Hawkes Process:} The $\alpha$ and $\omega$ gradients are clipped to $[0,10^3]$ with a learning rate of $1\times10^{-3}$. The $\beta$ line search is conducted over 20 equally-spaced points in a $\pm5\times10^{-3}$ radius surrounding the previous value of $\beta$. The optimization of $\alpha$ and $\omega$ terminates when the relative difference in validation NLL is less than $10^{-4}$. The algorithm terminates when the relative difference in validation NLL after line search is less than $10^{-4}$.
\end{itemize}
As before, $n_{\rm traj}=10$ trajectories are sampled from each model for each sensor. \rev{The fit parameters are displayed in Table~\ref{tab:fit_parameters}.}

\subsubsection{Results} The results can be found in Table~\ref{tab:prediction_window_metrics_outdoor} and Figure~\ref{fig:interval_stats_outdoor}. In this setting, all methods are less accurate. This can be attributed to the longer time scales and less controlled conditions in the outdoor environment. As demonstrated in Table~\ref{tab:prediction_window_metrics_outdoor}, the linear models poorly model the corrosion current, resulting in very small windows that are highly inaccurate, in general. The point process models achieve just over half the error of the linear models on the test dataset with reasonable window sizes.

\section{Discussion}\label{sec:discussion}
In this work, we have developed the novel notion of discrete degradation events that captures the degradation of a coating system, facilitating efficient modeling. We demonstrate that discrete events derived from continuous sensor measurements contain nearly as much information about degradation and failure as the continuous measurements themselves, while requiring orders of magnitude fewer measurements. This translates to a more efficient modeling framework, wherein a model of discrete events in continuous time is only required to accurately forecast $O(10)$ events, in contrast to a model of continuous sensor data required to accurately forecast $O(1000)$ measurements. In particular, we outline a marked Hawkes process to model and forecast degradation events, modeling self-influence and magnitude of events. 

These models are fit using each of the three event types on controlled laboratory and uncontrolled outdoor experiment data, given relatively few event sequences for effective model fitting. Compared with models of continuous sensor readings, the point process models enable more accurate prediction of failure with smaller uncertainties, also requiring fewer inputs. Modeling the degradation of various coating stackups enables a comparative analysis of their properties. Furthermore, characterizing the remaining protective life of a coated material permits need-based maintenance of coating systems, helping to reduce waste and needless reapplication of hazardous coating material while minimizing the risk of damage. An additional benefit of degradation event modeling is that it does not necessarily require embedded sensors on the coated materials. For instance, environment events can be verified using general weather-tracking databases. 

The primary limitations of this work are due to the small data setting, enabling limited model validation and evaluation. Laboratory and outdoor environment testing of coated materials is expensive, with experiments taking many months to complete. As such, we are limited to observations of degradation in only 30 coated panels. Additionally, comparing the efficacy of different modeling frameworks in forecasting degradation is non-trivial, as there is no straightforward, one-to-one comparison. Instead, we opted to compare discrete event modeling to continuous-time modeling with respect to a downstream task - failure prediction. In this way, we have demonstrated that modeling discrete events more accurately represent the degradation relevant to coating failure and blistering. Nonetheless, this study provides useful insight into the superior properties of discrete event modeling to forecast degradation.

\section*{Acknowledgement}
{We would like to extend gratitude to Brandi Clark, Victoria Avance, and Rebecca Marshall for their considerable contributions to this work, including providing valuable expertise in materials science and engineering. We would also like to thank Lasya Akshara for her contribution to the data science. This material is based upon work supported wholly or in part by the United States Army Corps of Engineers through the Strategic Environmental Research and Development Program (SERDP) under Contract No W912HQ19C0062. Any opinions, findings and conclusions or recommendations expressed in this material are those of the author(s) and do not necessarily reflect the views of the United States Army Corps of Engineers or the Strategic Environmental Research and Development Program (SERDP).}

\bibliographystyle{plain} 
\bibliography{main}

\end{document}